%% file: main.tex
\documentclass{webofc}
\usepackage[varg]{txfonts}   
\usepackage{subfigure}
\usepackage{booktabs}
\usepackage{multirow}

\newcommand{\eg}{\textit{e.g.} }
\newcommand{\ie}{\textit{i.e.} }

\begin{document}
\title{A low-mass galaxy cluster as a test-case study for the NIKA2 SZ Large Program}

\author{\firstname{F.} \lastname{K\'eruzor\'e} \inst{\ref{LPSC}}
        \fnsep\thanks{\email{keruzore@lpsc.in2p3.fr}} 
\and \firstname{R.} \lastname{Adam} \inst{\ref{LLR},\ref{CEFCA}}
\and \firstname{P.} \lastname{Ade} \inst{\ref{Cardiff}}
\and \firstname{P.} \lastname{Andr\'e} \inst{\ref{CEA1}}
\and \firstname{A.} \lastname{Andrianasolo} \inst{\ref{IPAG}}
\and \firstname{M.} \lastname{Arnaud} \inst{\ref{CEA1}}
\and \firstname{H.} \lastname{Aussel} \inst{\ref{CEA1}}
\and \firstname{I.} \lastname{Bartalucci} \inst{\ref{CEA1}}
\and \firstname{A.} \lastname{Beelen} \inst{\ref{IAS}}
\and \firstname{A.} \lastname{Beno\^it} \inst{\ref{Neel}}
\and \firstname{A.} \lastname{Bideaud} \inst{\ref{Neel}}
\and \firstname{O.} \lastname{Bourrion} \inst{\ref{LPSC}}
\and \firstname{M.} \lastname{Calvo} \inst{\ref{Neel}}
\and \firstname{A.} \lastname{Catalano} \inst{\ref{LPSC}}
\and \firstname{B.} \lastname{Comis} \inst{\ref{LPSC}}
\and \firstname{M.} \lastname{De~Petris} \inst{\ref{Roma}}
\and \firstname{F.-X.} \lastname{D\'esert} \inst{\ref{IPAG}}
\and \firstname{S.} \lastname{Doyle} \inst{\ref{Cardiff}}
\and \firstname{E.F.C.} \lastname{Driessen} \inst{\ref{IRAMF}}
\and \firstname{A.} \lastname{Gomez} \inst{\ref{CAB}}
\and \firstname{J.} \lastname{Goupy} \inst{\ref{Neel}}
\and \firstname{C.} \lastname{Kramer} \inst{\ref{IRAME}}
\and \firstname{B.} \lastname{Ladjelate} \inst{\ref{IRAME}}
\and \firstname{G.} \lastname{Lagache} \inst{\ref{LAM}}
\and \firstname{S.} \lastname{Leclercq} \inst{\ref{IRAMF}}
\and \firstname{J.-F.} \lastname{Lestrade} \inst{\ref{LERMA}}
\and \firstname{J.F.} \lastname{Mac\'ias-P\'erez} \inst{\ref{LPSC}}
\and \firstname{P.} \lastname{Mauskopf} \inst{\ref{Cardiff},\ref{Arizona}}
\and \firstname{F.} \lastname{Mayet} \inst{\ref{LPSC}}
\and \firstname{A.} \lastname{Monfardini} \inst{\ref{Neel}}
\and \firstname{L.} \lastname{Perotto} \inst{\ref{LPSC}}
\and \firstname{G.} \lastname{Pisano} \inst{\ref{Cardiff}}
\and \firstname{E.} \lastname{Pointecouteau} \inst{\ref{IRAP}}
\and \firstname{N.} \lastname{Ponthieu} \inst{\ref{IPAG}}
\and \firstname{G.W.} \lastname{Pratt} \inst{\ref{CEA1}}
\and \firstname{V.} \lastname{Rev\'eret} \inst{\ref{CEA1}}
\and \firstname{A.} \lastname{Ritacco} \inst{\ref{IRAME}}
\and \firstname{C.} \lastname{Romero} \inst{\ref{IRAMF}}
\and \firstname{H.} \lastname{Roussel} \inst{\ref{IAP}}
\and \firstname{F.} \lastname{Ruppin} \inst{\ref{MIT}}
\and \firstname{K.} \lastname{Schuster} \inst{\ref{IRAMF}}
\and \firstname{S.} \lastname{Shu} \inst{\ref{IRAMF}}
\and \firstname{A.} \lastname{Sievers} \inst{\ref{IRAME}}
\and \firstname{C.} \lastname{Tucker} \inst{\ref{Cardiff}}
\and \firstname{R.} \lastname{Zylka} \inst{\ref{IRAMF}}}

\institute{
  \label{LPSC} Univ. Grenoble Alpes, CNRS, Grenoble INP, LPSC-IN2P3, 53, avenue des Martyrs, 38000 Grenoble, France
  \and \label{LLR} LLR (Laboratoire Leprince-Ringuet), CNRS, École Polytechnique, Institut Polytechnique de Paris, Palaiseau, France  
  \and \label{CEFCA} Centro de Estudios de Física del Cosmos de Aragón (CEFCA), Plaza San Juan, 1, planta 2, E-44001, Teruel, Spain 
  \and \label{Cardiff} Astronomy Instrumentation Group, University of Cardiff, UK          
  \and \label{CEA1} AIM, CEA, CNRS, Université Paris-Saclay, Université Paris Diderot, Sorbonne Paris Cité, 91191 Gif-sur-Yvette, France    
  \and \label{IPAG} Univ. Grenoble Alpes, CNRS, IPAG, 38000 Grenoble, France     
  \and \label{IAS} Institut d'Astrophysique Spatiale (IAS), CNRS and Universit\'e Paris Sud, Orsay, France    
  \and \label{Neel} Institut N\'eel, CNRS and Universit\'e Grenoble Alpes, France
  \and \label{Roma} Dipartimento di Fisica, Sapienza Universit\`a di Roma, Piazzale Aldo Moro 5, I-00185 Roma, Italy       
  \and \label{IRAMF} Institut de RadioAstronomie Millim\'etrique (IRAM), Grenoble, France 
  \and \label{CAB} Centro de Astrobiolog\'ia (CSIC-INTA), Torrej\'on de Ardoz, 28850 Madrid, Spain
  \and \label{IRAME} Instituto de Radioastronom\'ia Milim\'etrica (IRAM), Granada, Spain 
  \and \label{LAM} Aix Marseille Univ, CNRS, CNES, LAM (Laboratoire d’Astrophysique de Marseille), Marseille, France
  \and \label{LERMA} LERMA, Observatoire de Paris, PSL Research University, CNRS, Sorbonne Universités, UPMC Univ. Paris 06, 75014 Paris,
  France
  \and \label{Arizona} School of Earth and Space Exploration and Department of Physics, Arizona State University, Tempe, AZ 85287         
  \and \label{IRAP} IRAP, Université de Toulouse, CNRS, CNES, UPS, (Toulouse), France
  \and \label{IAP} Institut d’Astrophysique de Paris, CNRS (UMR7095), 98 bis boulevard Arago, 75014 Paris, France
  \and \label{MIT} Kavli Institute for Astrophysics and Space Research, Massachusetts Institute of Technology, Cambridge, MA 02139, USA 
  }

\abstract{%
  High-resolution mapping of the hot gas in galaxy clusters is a key tool for cluster-based cosmological analyses. 
  Taking advantage of the NIKA2 millimeter camera operated at the IRAM 30-m telescope, the NIKA2 SZ Large Program seeks to get a high-resolution follow-up of 45 galaxy clusters covering a wide mass range at high redshift in order to re-calibrate some of the tools needed for the cosmological exploitation of SZ surveys. 
  We present the second cluster analysis of this program, targeting one of the faintest sources of the sample in order to tackle the difficulties in data reduction for such faint, low-SNR clusters.
  In this study, the main challenge is the precise estimation of the contamination by sub-millimetric point sources, which greatly affects the tSZ map of the cluster. 
  We account for this contamination by performing a joint fit of the SZ signal and of the flux density of the compact sources.
  A prior knowledge of these fluxes is given by the adjustment of the SED of each source using data from both NIKA2 and the \textit{Herschel} satellite.
  The first results are very promising and demonstrate the possibility to estimate thermodynamic properties with NIKA2, even in a compact cluster heavily contaminated by point sources.
  }
\maketitle
\section{Introduction}\label{sec_intro}
Galaxy clusters have been shown to be excellent probes of large-scale structure formation processes and of the underlying cosmology \cite{bocquet_cluster_2019, planck_collaboration_planck_2016-1}.
Recent CMB experiments have provided us with large catalogs of galaxy clusters (\eg \cite{planck_collaboration_planck_2016, bleem_galaxy_2015, hilton_atacama_2018}) detected through their imprint on the CMB by the thermal Sunyaev-Zel'dovich effect (tSZ, \cite{sunyaev_observations_1972}).
The analysis of these catalogs have revealed a $\sim\hspace{-4pt}2\sigmaup$ discrepancy between cosmological parameters obtained from galaxy clusters and from CMB anisotropies \cite{planck_collaboration_planck_2018, salvati_constraints_2018}.
Possible explanations for this tension include new physics (\eg neutrino masses \cite{bolliet_including_2019} or a discrepancy between local and distant Universe probes \cite{wong_h0licow_2019}), or uncontrolled systematics in the analysis, such as a bias in the mean pressure profile of galaxy clusters, the tSZ--mass scaling relation, or the hydrostatic mass bias (see \eg \cite{ruppin_impact_2019, salvati_mass_2019}).

The NIKA2 SZ Large Program (LPSZ, \cite{mayet_cluster_2019}) seeks to investigate these possible systematics through an evaluation of the mean pressure profile and the SZ--mass scaling relation at high redshift and with a high angular resolution. 
It takes advantage of the NIKA2 camera performance \cite{adam_nika2_2018, calvo_nika2_2016, perotto_calibration_2019}, \ie its capabilities of in-depth mapping of a large field of view at a high angular resolution simultaneously in 2 frequency bands. 
These characteristics are particularly well-suited for SZ observations.
The first cluster analysis within the LPSZ \cite{ruppin_first_2018} was a science verification study that showed the NIKA2 ability to produce excellent SZ maps of a massive, intermediate redshift cluster with a large observation time.
For the second analysis, we focus on a fainter target, in order to evaluate the quality of the SZ data that can be expected from NIKA2 for clusters on the lower end of the mass range of the LPSZ sample.

\section{The ACT-CL J0215.4+0030 galaxy cluster}\label{sec:data}

We chose to analyze the ACT-CL J0215.4+0030 galaxy cluster, which is a low mass and high redshift source.
It was first detected by the Atacama Cosmology Telescope (ACT, \cite{hilton_atacama_2018}), and was not detected by the \textit{Planck} match-filtering. 
Therefore, it doesn't have any counterpart in the \textit{Planck}  SZ catalogs.
In addition, the cluster was observed by XMM-\textit{Newton} for 34 ks, which is sufficient to estimate a temperature profile from X-ray spectroscopy.
This will allow us to compare the thermodynamic profiles obtained by combining SZ and X-ray data (pressure and density) to those obtained with X-rays only (density and temperature).
The main characteristics of this cluster are shown in Table \ref{tab:pszvact}. 
They highlight the faintness of this source compared to the cluster used for the first analysis of the LPSZ sample \cite{ruppin_first_2018}.
\begin{table}
  \hfill
\begin{minipage}[c]{0.5\textwidth}
  \footnotesize
  \begin{tabular}{c c c}
    \toprule
      &  PSZ2-G144.83 &  ACT-CL J0215.4 \\
      &  +25.11  &  +0030 \\
    \midrule\midrule
    $z$  &  0.584  &  0.865 \\
    \midrule
    \multirow{2}{*}{$M_{500}$}  &  $8.2\times 10^{14}\,\mathrm{M}_\odot$   &  $3.8\times 10^{14}\,\mathrm{M}_\odot$  \\
    &  (\textit{Planck})  &  (ACT)  \\
    \midrule
    $\theta_{500}$  &  2.9 arcmin  &  1.6 arcmin \\
    \midrule
    $t_\mathrm{obs}/t_\textsc{lpsz}$  &  11h / 2h = 5.5  &  9h / 9h = 1 \\
    \midrule
    tSZ decrement  &  \multirow{2}{*}{13.5$\sigmaup$}  &  \multirow{2}{*}{8.5$\sigmaup$} \\
    peak  &  &  \\
    \bottomrule
  \end{tabular}
\end{minipage}
\hfill
\begin{minipage}[c]{0.42\textwidth}
  \normalsize
  \vspace{10pt}
  \caption{%
          Comparison between the first analyzed cluster of the LPSZ (PSZ2-G144, \cite{ruppin_first_2018}) and our target (ACT-CL J0215). %
          We note large differences in mass, redshift and observed to requested time ratio, meaning that we expect the signal-to-noise ratio in the ACT-CL~J0215 maps to be smaller than what was obtained for PSZ2-G144.
          }
  \label{tab:pszvact}
\end{minipage}
\hfill
\end{table}

The cluster was observed during 9 hours, \ie its planned LPSZ time, in January 2018, with an average zenith atmospheric opacity at 225 GHz of $0.175$ and an average elevation of $43^\circ$, which are standard conditions for winter observations at the IRAM 30 m telescope.

\begin{figure}[h]
  \hfill
  \subfigure{\includegraphics[height=4.8cm]{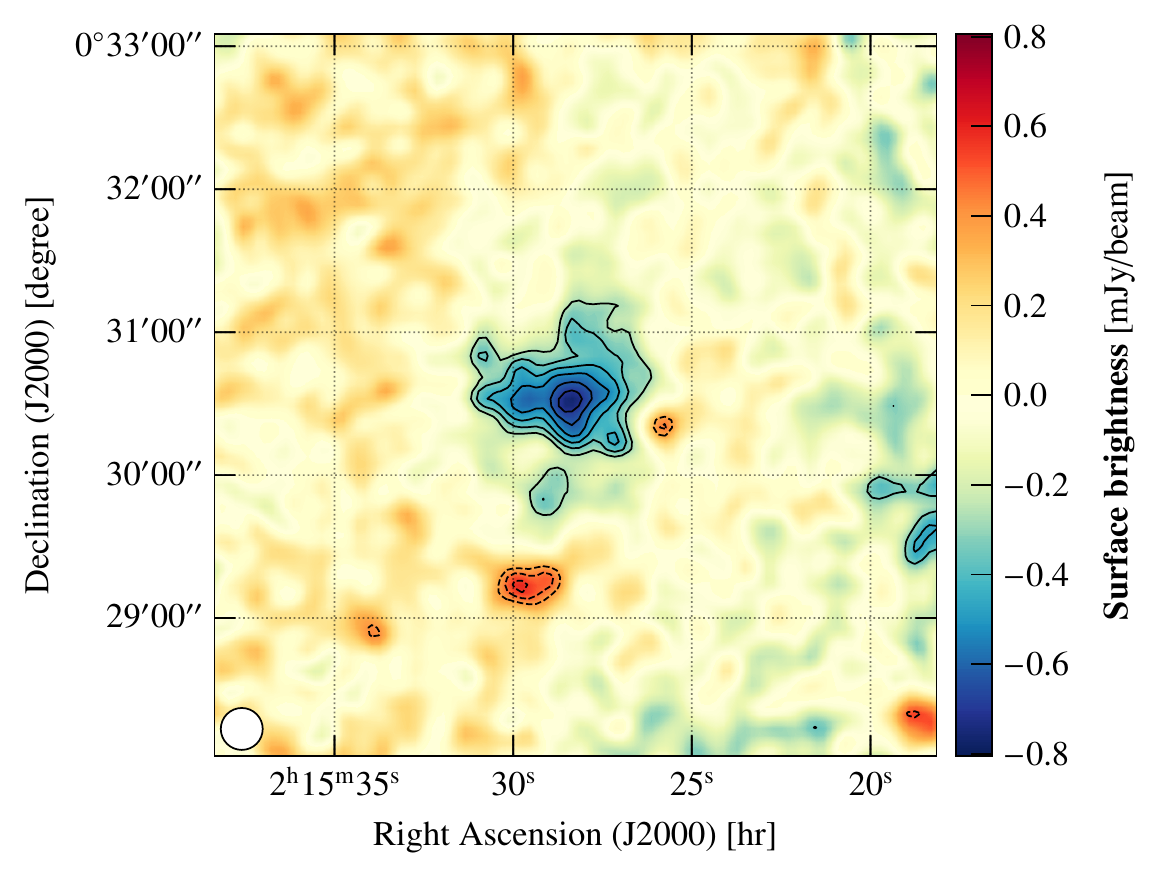}}
  \hfill
  \subfigure{\includegraphics[height=4.8cm]{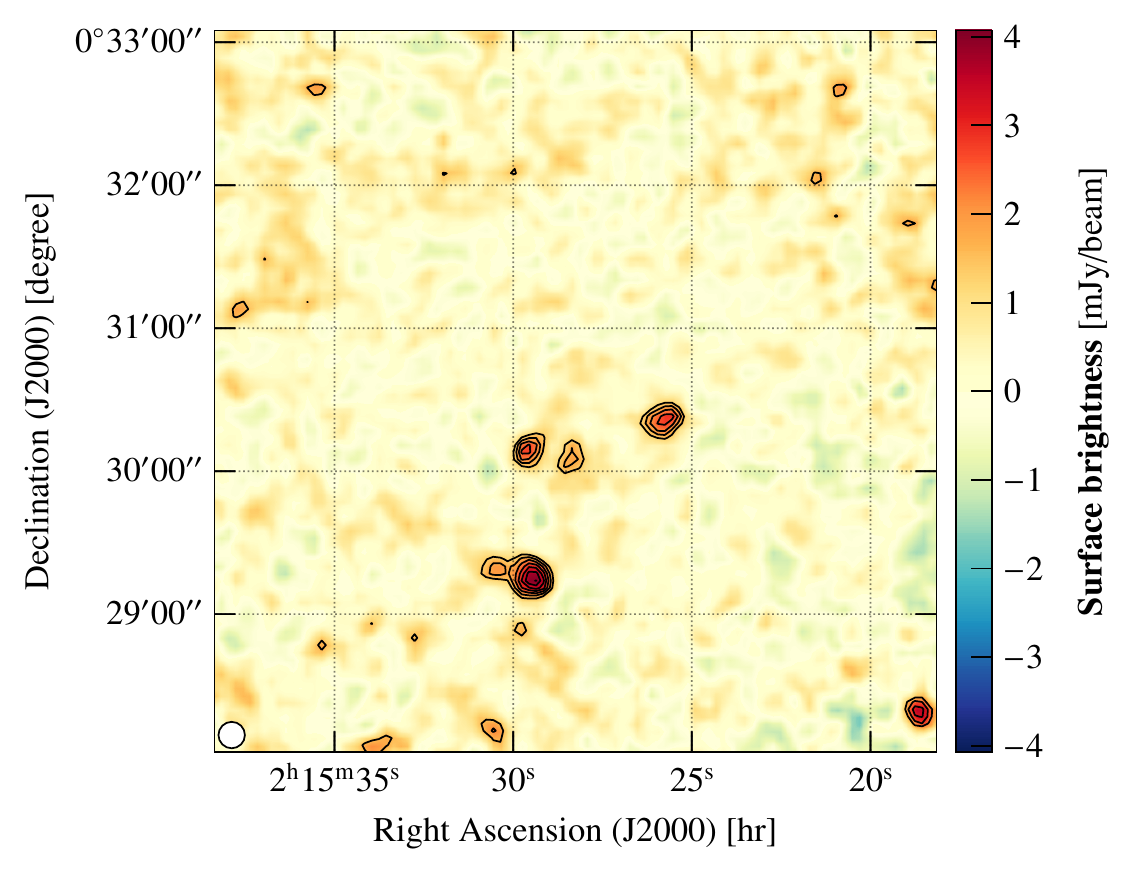}}
  \hfill
  \caption{%
           NIKA2 maps of the cluster in the 150~GHz (\textit{left}) and 260~GHz (\textit{right}) bandpasses. %
           Both maps are shown in a 5'$\times$5' area centered on the observed coordinates. %
           The NIKA2 instrumental beam's FWHM is represented as a white disk in the bottom left corner of each map. %
           Contours show SNR levels starting from $\pm 3\sigmaup$ with a $1\sigmaup$ spacing. %
           The left (right) map is smoothed with a 10'' (6'') gaussian kernel for display purposes.
          }
  \label{fig:nk2maps}
\end{figure}

The NIKA2 data are reduced and calibrated using the baseline calibration method as described in \cite{perotto_calibration_2019}. For the noise decorrelation we use the \textit{Most Correlated Pixels} method, where a common mode is estimated on groups of most correlated detectors and subtracted from the time-ordered data \cite{perotto_calibration_2019}. 
The resulting maps are presented in Figure \ref{fig:nk2maps}.
In the 150~GHz map (left panel), we identify the cluster as a decrement in the center of the map, detected at a peak SNR of $\sim 8\sigmaup$.
We also note in this map the presence of large residual noise structures surrounding the cluster, as well as positive point sources, some of them being close enough to affect our signal (within $R_{500}$).
In the 260~GHz map (right), we do not expect to detect any SZ signal given the shape of the SZ spectrum and the NIKA2 sensitivity \cite{perotto_calibration_2019}.
Therefore, we observe a seemingly empty field with the positive point sources mentioned above.

\section{Contamination by sub-millimetric point sources}\label{sec:ps}

The shape of the tSZ effect spectrum and NIKA2 sensitivity at 260~GHz prevent us from detecting any SZ signal in this band given the considered observing time.
Therefore, sub-millimetric point sources can be easily identified in the NIKA2 260~GHz map because they are expected to be the dominant component of the signal in this band.
Some of these sources are very close to the cluster, and can therefore affect the reconstruction of its shape, since a positive flux can compensate the SZ decrement and create ``holes'' in the cluster map.
Given the faintness of the SZ effect for this cluster (less than a mJy at its peak), its small size, and the strong fluxes of the considered point sources (several mJy at 260~GHz), we expect these sources to have a strong effect on our SZ map, making the estimation of this contamination in the NIKA2 150~GHz map the main challenge of this analysis.
We identify four sources near the cluster in the NIKA2 260~GHz map, which are identified as sub-millimetric sources by coordinates cross-matching in the \textit{Herschel} Stripe 82 Survey (HerS, \cite{viero_herschel_2014}). 
A fifth source in the north-eastern region of the cluster with a SNR lower than 3 at 260 GHz is also identified.

\begin{figure}[t]
  \hfill
  \subfigure{\includegraphics[height=4.8cm]{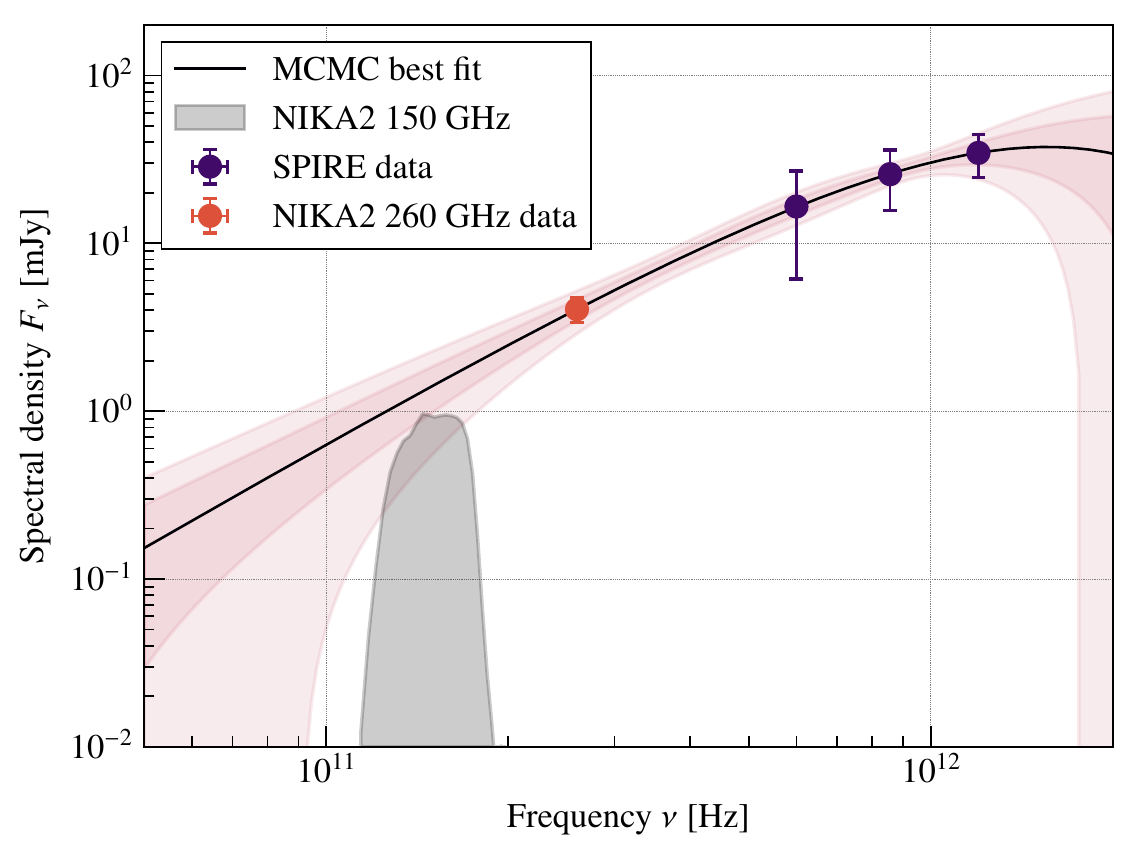}}
  \hfill
  \subfigure{\includegraphics[height=4.8cm]{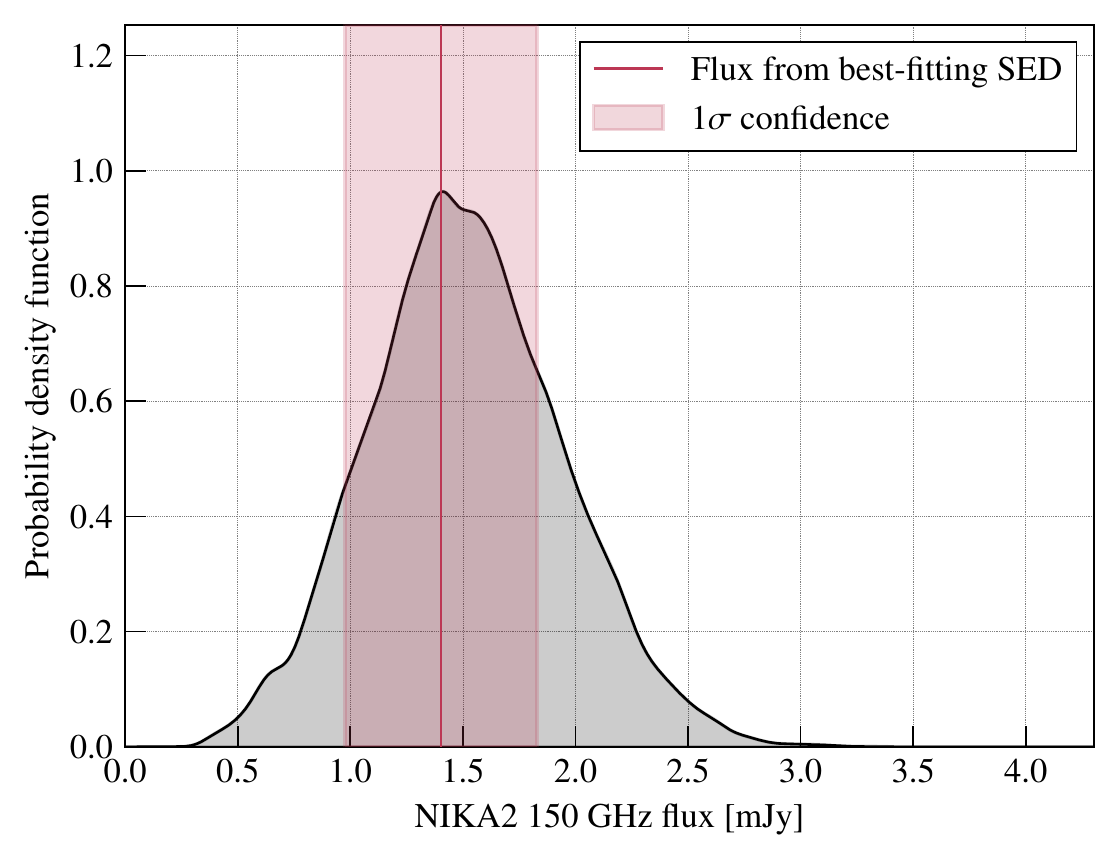}}
  \hfill
  \caption{%
           Results of the estimation of the contamination in the NIKA2 150~GHz map for one of the sources in our field. %
           \textit{Left:} SED results.
           The NIKA2 150~GHz point is extrapolated from the fit by integrating the SED distribution in the NIKA2 150~GHz bandpass. %
           The black line marks the best-fitting SED, and the enveloppes show the 1 and 2$\sigmaup$ confidence intervals. %
           \textit{Right:} inferred probability distribution for the 150~GHz flux of the same source. %
           The vertical line shows the flux computed from the best-fitting SED.
          }
  \label{fig:psresults}
\end{figure}

The contamination of the NIKA2 150~GHz map by each sub-millimetric point-source can be estimated by fitting its spectral energy distribution (SED).
To do so, we need to know the flux of each source at several frequencies.
We take their fluxes in each band of the SPIRE instrument, \ie 250, 350 and 500 $\muup\mathrm{m}$ (1200, 860 and 600 GHz respectively) in the HerS catalog \cite{viero_herschel_2014}.
To add to these three sub-mm frequencies, we fit each source in the NIKA2 260~GHz map as a sum of three 2D Gaussian functions, which was found to be an accurate description of the NIKA2 instrumental beam \cite{perotto_calibration_2019}.
The overall amplitude of this fit gives us the 260~GHz flux of the source.
Then, MCMC sampling is used to fit the SED of each source with a grey-body spectrum.
For each sample of the posterior distribution of parameters, a SED is computed and integrated in the NIKA2 150~GHz bandpass. 
This gives us a sampled distribution of 150~GHz fluxes, which is used to compute an estimation of the probability density of the flux of the source in the NIKA2 map through a kernel density estimation.
The results of the SED fit and of the NIKA2 150~GHz flux extrapolation for the source in the western part of the cluster are given in the left and right panels of Figure \ref{fig:psresults}, respectively.

\section{SZ model adjustment}\label{sec:results}

The tSZ surface brightness map observed in the NIKA2 150~GHz band is directly proportional to the Compton parameter $y$, in turn proportional to the electron pressure of the intra-cluster medium integrated along the line of sight.
Therefore, assuming a spherical symmetry, a pressure profile of the galaxy cluster can be fitted on the NIKA2 SZ map, along with a nuisance parameter for the calibration coefficient.
We do so by using the NIKA2 LPSZ pipeline \cite{ruppin_cosmologie_2018}. 

The point-source contamination needs to be accounted for in order to retrieve accurate profiles.
Rather than merely estimating the contamination and subtracting them from the map, we take them into account by including them in the fitting procedure. 
Five parameters are added to the MCMC sampling, corresponding to the fluxes of the five sub-millimetric point sources, their position in the map being precisely known from the adjustment in the NIKA2 260~GHz map.
Our prior knowledge of these five fluxes is given by the probability distribution that we have computed from the SED fitting described in Section \ref{sec:ps}.
This method allows us to take full advantage of the outputs of the SED fit results, \ie a probability distribution for the flux of each source rather than a single value and uncertainty.

\begin{figure}[t]
  \centering
  \includegraphics[width=.95\linewidth]{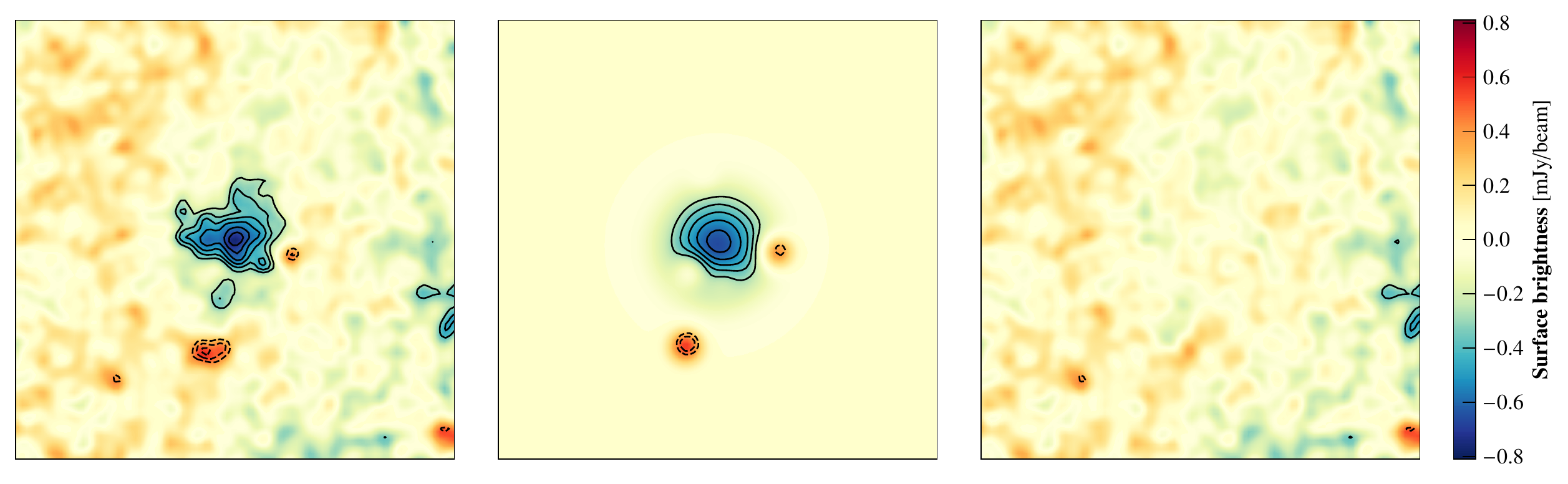}
  \caption{%
           Results of the fit of the 150~GHz map of ACT-CL J0215.4+0030. %
           From left to right, we present the data, the best-fitting model (including the Gaussian models of accounted point sources), and the residuals. %
           Each map is given in the same 5'$\times$5' area with the same color scale and smoothed with a 10'' gaussian kernel for display purposes. %
           In each map, the contours give the SNR levels starting from $\pm 3\sigmaup$ with a $1\sigmaup$ spacing.
          }
  \label{fig:panco}
\end{figure}

The results of this MCMC analysis considering a generalized Navarro-Frenk-and-White pressure profile (gNFW, \cite{nagai_effects_2007}) are shown in Figure \ref{fig:panco}. 
From left to right, we present the NIKA2 150~GHz data, best-fitting model for the cluster and point sources, and residuals.
No high-SNR residuals can be identified, indicating no significant difference between the data and description of the map as a spherical gNFW cluster with point sources.
This comforts us in our assumption of spherical gNFW profile, as well as in the quality of our estimation of the contamination by point sources.

As mentioned previously, the NIKA2 pressure profile estimate resulting from this analysis can then be compared to XMM-\textit{Newton} results.
The results of NIKA2+XMM-imaging and the XMM-only profiles are found to be consistent, within 1$\sigmaup$ from one another.
More details on the analysis and the recovered profiles and integrated quantities will be given in \cite{keruzore_act-cl_2019}.

\section{Conclusions}\label{sec:conclu} 
The analysis of the ACT-CL J0215.4+0030 galaxy cluster has revealed the challenges associated to faint galaxy clusters in the NIKA2 SZ Large Program.
First, correlated noise structures remain in our maps after the decorrelation process, indicating that some progress in this regard could be beneficial.
Second, we saw how important the contamination by point sources can be for such faint tSZ targets.
In this particular case, strong sub-millimetric point sources in the peripheral region of the cluster have a large impact on the global shape of the cluster, and can therefore greatly affect the pressure profile and thermodynamic properties of the cluster if they are not carefully accounted for in the analysis.
NIKA2 proved to be highly efficient in this regard, its dual-band capacities and high angular resolution allowing us to detect sources and measure their flux in the 260~GHz band.
However, we need to rely on external \textit{Herschel} data for the SED adjustments, raising the question of possible biases in the final cosmological products of the LPSZ -- \eg mean pressure profile or SZ--mass scaling relation -- due to contamination by undocumented sources.

\section*{Acknowledgements}
\input{acknowledgements}

\bibliography{Proceeding_mmUniverse}

\end{document}

%% file: acknowledgements.tex
We would like to thank the IRAM staff for their support during the campaigns. The NIKA dilution cryostat has been designed and built at the Institut N\'eel.
In particular, we acknowledge the crucial contribution of the Cryogenics Group, and in particular Gregory Garde, Henri Rodenas, Jean Paul Leggeri, Philippe Camus. This work has been partially funded by the Foundation Nanoscience Grenoble and the LabEx FOCUS ANR-11-LABX-0013.
This work is supported by the French National Research Agency under the contracts "MKIDS", "NIKA" and ANR-15-CE31-0017 and in the framework of the "Investissements d’avenir” program (ANR-15-IDEX-02).
This work has benefited from the support of the European Research Council Advanced Grants ORISTARS and M2C under the European Union's Seventh Framework Programme (Grant Agreement Nos. 291294 and 340519).
We acknowledge fundings from the ENIGMASS French LabEx (R. A. and F. R.), the CNES post-doctoral fellowship program (R. A.), the CNES doctoral fellowship program (A. R.) and the FOCUS French LabEx doctoral fellowship program (A. R.).
R.A. acknowledges support from Spanish Ministerio de Econom\'ia and Competitividad (MINECO) through grant number AYA2015-66211-C2-2.